\documentclass[aip,reprint]{revtex4-1}
\usepackage{amsmath}
\usepackage{amsfonts}
\usepackage{amssymb}
\usepackage{graphicx}
\usepackage{braket}
\usepackage{comment}
\usepackage{subcaption}
\usepackage{ulem}
\usepackage{xcolor}

\usepackage[hidelinks]{hyperref}
\draft 

\begin{document}

\title{High-brightness fiber-based Sagnac source of entangled photon pairs for multiplexed quantum networks} 

\author{Tess Troisi}
\author{Yoann Pelet}
\author{Romain Dalidet}
\author{Gregory Sauder}
\author{Olivier Alibart}
\author{Sébastien Tanzilli}
\author{Anthony Martin}
\affiliation{Institut de Physique de Nice, UMR 7010, Université Côte d'Azur, CNRS}

\date{\today}

\begin{abstract}
A fully fibered source of entangled photon pairs based on a nonlinear Sagnac interferometer is reported. Operating at telecom wavelengths, the source relies exclusively on standard fiber-optic components and periodically poled lithium niobate (PPLN) waveguides, resulting in a compact, robust, and field-deployable architecture. The generation stage supports both polarization and energy-time entanglement without modification, enabling versatile operation depending on the targeted application.
Broadband spontaneous parametric down-conversion allows dense wavelength-division multiplexing over the telecom C and L bands. High normalized brightness (10.3 kpairs/s/nm/mW$^2$) is achieved on a standard 100\,GHz ITU channel pair, together with high entanglement quality. Polarization and energy-time encodings are characterized through state tomography and two-photon interference measurements, yielding fidelities, purities, and visibilities exceeding 96\% over multiple wavelength channels.
The stability and reproducibility of the source are further evaluated through long-duration operation in a network environment. These results demonstrate that the proposed Sagnac source constitutes a practical and scalable building block for future plug-and-play quantum communication and quantum networking platforms.
\end{abstract}

\pacs{}

\maketitle 
\section{Introduction}

Photonic entanglement is a key resource for quantum communication, distributed quantum information processing, and emerging quantum network architectures. In particular, quantum key distribution (QKD) has reached a level of maturity that enables deployment beyond laboratory environments toward real-field implementations~\cite{Sena:25,Wengerowsky_2018,wehner2018}. Practical quantum communication systems, therefore, require sources that combine high entanglement quality with robustness, ease of use, and compatibility with existing fiber-optic infrastructure.

Among the various photonic encodings, energy-time entanglement is well suited for long-distance fiber transmission and high-dimensional encoding~\cite{Xavier2025}, while polarization encoding offers advantages in terms of state manipulation and measurement, and remains the preferred choice for free-space links, with efficient fiber-based implementations also demonstrated~\cite{Qotham}. The development of versatile entangled photon sources capable of supporting multiple encodings is therefore an important step toward flexible quantum communication architectures.

From an engineering perspective, relevant performance metrics include high source brightness with high entanglement quality, low power consumption, robustness to environmental perturbations, ease of use, and straightforward integration with standard telecom components. While integrated photonic platforms provide compactness and excellent interferometric stability~\cite{Zhu_25}, coupling losses at the chip–fiber interface can limit their overall efficiency. Guided-wave and fiber-based approaches, therefore, remain attractive alternatives, offering bright, compact, and versatile entangled photon sources supporting multiple encoding schemes~\cite{Achatz2023,Gianini2026,Vergyris_2019}.
Taken together, these requirements are closely aligned with the long-term vision of large-scale quantum networks, in which stable, interoperable, and deployable entanglement sources play a central role~\cite{wehner2018}.

In this work, a fiber-based Sagnac entangled photon-pair source operating at telecom wavelengths is presented. The source is designed as a versatile and transportable device supporting both polarization and energy-time entanglement, without modification of the generation stage. Energy-time entanglement is provided by spontaneous parametric down-conversion, while polarization entanglement is generated using a polarization-dependent interferometric configuration. The broadband emission of the source enables dense wavelength-division multiplexing \cite{Arahira:11,Li_2023,Cabrejo_Ponce_2022}, and its intrinsic stability allows long-term operation without realignment.

The performance of the source is assessed through detailed polarization-entanglement characterization across multiple wavelength channel pairs. In contrast, energy-time encoding is evaluated through a quantum key distribution (QKD) experiment performed on a single channel pair, providing a system-level validation of the source stability under extended operation\cite{yoyo}.

\section{Nonlinear Sagnac source}

Sagnac interferometers have long been employed in precision sensing and metrology, most notably in fiber-optic gyroscopes, where their passive phase stability and robustness have been extensively exploited~\cite{Arditty1981}. These characteristics further motivate the use of Sagnac-based architectures for quantum technologies requiring long-term stability and resilience to environmental perturbations. Sagnac interferometers based on spontaneous parametric down-conversion (SPDC) have proven to be a particularly attractive platform for entangled photon-pair generation \cite{shi2004generation,kim2006phase,Kuzucu_2008}. Their intrinsic phase stability, arising from the counter-propagation of optical fields along identical paths, enables the generation of high-quality entangled states without active stabilization \cite{kim2001}.

The entangled photon-pair source is based on a nonlinear Sagnac interferometer~\cite{shi2004generation,kim2006phase}, conceptually depicted in \figurename~\ref{fig:sagnac_scheme}. A continuous-wave telecom pump laser is injected into the interferometer through a circulator and split into two counter-propagating paths by a polarizing beam splitter (PBS). The clockwise (CW) and counterclockwise (CCW) components propagate along identical optical paths with orthogonal polarizations.

In each propagation direction, the pump field undergoes second-harmonic generation (SHG) in a PPLN waveguide crystal. The generated visible field then pumps a second PPLN waveguide, where broadband degenerate photon pairs are produced via type-0 SPDC. The photon pairs generated in the CW and CCW contributions are coherently recombined at the polarizing beam splitter, resulting in a polarization-entangled state of the form
\begin{equation}
\ket{\Psi} = \alpha \ket{HH} + e^{i\phi}\beta \ket{VV},
\end{equation}
where $|\alpha|^2 + |\beta|^2 = 1$. The relative amplitudes $\alpha$, $\beta$ and phase $\phi$ are controlled via the pump's input polarization.

The fully fibered implementation of the Sagnac interferometer, combined with the use of standard telecom components, results in a compact and mechanically stable source. The broadband phase-matching of the PPLN waveguides enables photon-pair generation over a spectral width of approximately 90~nm (FWHM), allowing efficient wavelength multiplexing across multiple standard ITU channels.

\section{Experimental setup}
\begin{figure}[ht]
    \centering
    \includegraphics[width=1\linewidth]{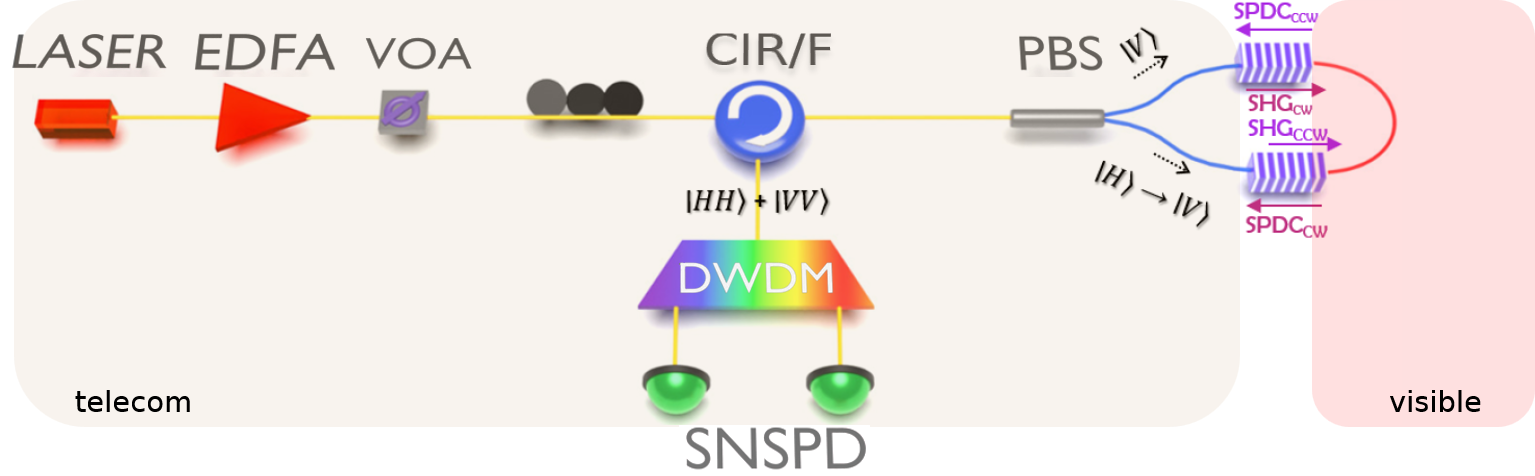}
    \caption{Experimental setup of the fiber-based Sagnac source. EDFA: erbium-doped fiber amplifier; VOA: variable optical attenuator; PC: polarization controller; CIR: circulator;
PBS: polarizing beam splitter; NF: notch filter; DWDM: dense wavelength-division
multiplexer.}
    \label{fig:sagnac_scheme}
\end{figure}

\begin{figure}[ht]
\centering
\includegraphics[width=1\linewidth]{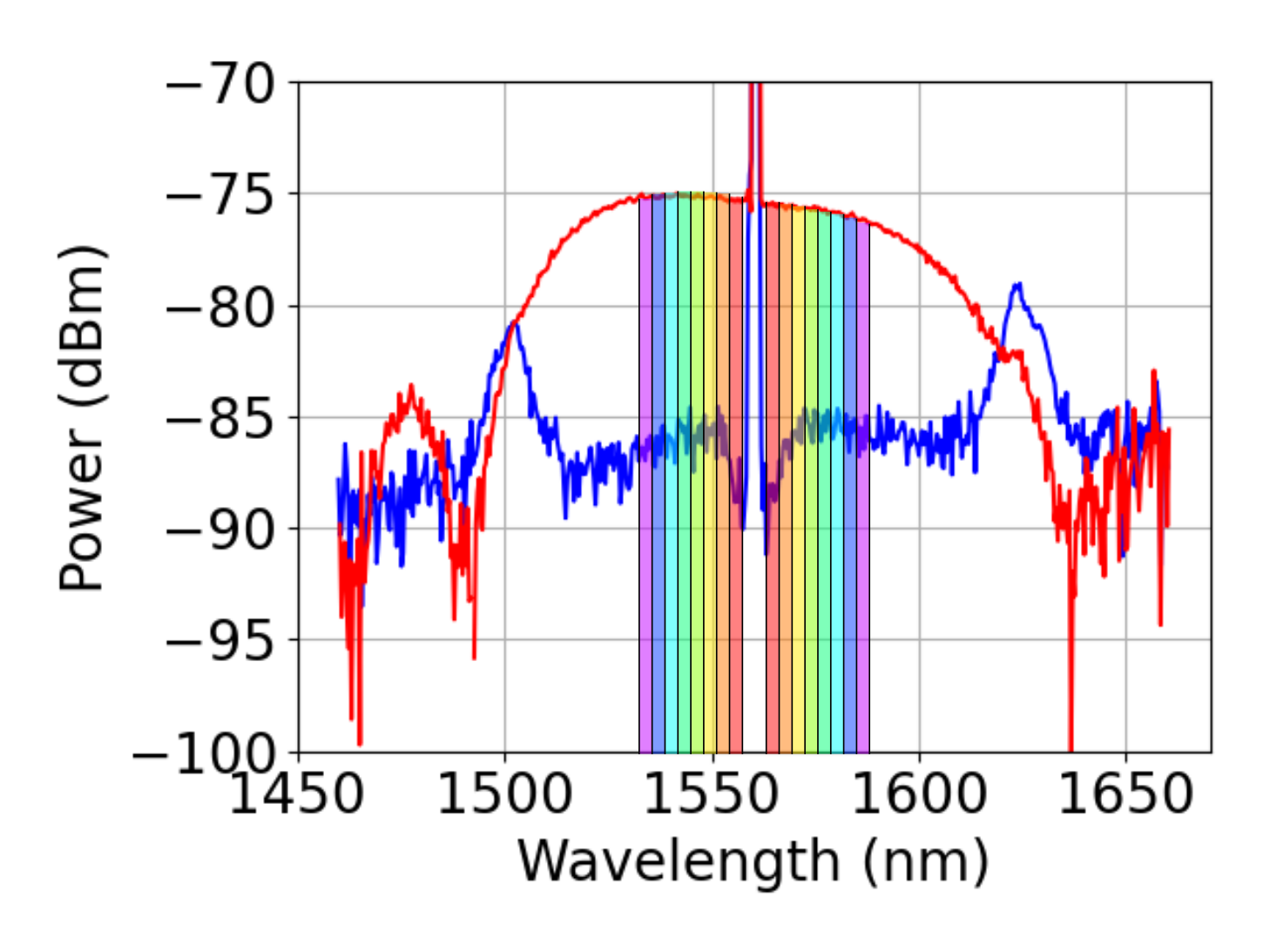}
    \caption{Comparison of the photon pair spectrum (red) with the spontaneous Raman scattering spectrum (blue). Exploited photon pair spectrum for entanglement distribution over 100\,GHz DWDM management plan is represented here (not to scale).}
    \label{fig:raman}
\end{figure}

As depicted in \figurename~\ref{fig:sagnac_scheme}, a continuous-wave laser centered at 1560.6\,nm with a 1\,kHz linewidth is used as the pump. It is further amplified using an erbium-doped fiber amplifier (EDFA), and its power is finely controlled by a variable optical attenuator (VOA) before injection into the Sagnac interferometer. A fiber circulator, including a notch filter, suppresses the laser amplified spontaneous emission and forward Raman scattering, ensuring that only backward Raman scattering noise contributes to the detected spectrum, as shown in \figurename~\ref{fig:raman}.
The polarization state of the pump mode is achieved using a thermal polarization controller before the PBS, to define the power splitting ratio and the phase relation between the CW and CCW propagations. The output ports of the PBS are aligned to the slow axes of polarization-maintaining fibers to ensure proper coupling into the nonlinear waveguides.
When the pump polarization is set to a $\frac{\ket{H} + \ket{V}}{\sqrt{2}}$, after the double conversion processes, the desired state $\ket{HH}+\ket{VV}/\sqrt{2}$ is obtained. 
The conversion processes are ensured by type-0 PPLN:MgO fiber-pigtailed waveguide crystals with a length of 22\,mm. Careful selection of the PPLN nonlinear crystals has been applied in order to obtain performances as similar as possible. The SHG conversion efficiencies and SPDC characteristics of the PPLN waveguides 1 and 2 are summarized in  TABLE~\ref{tab:my_label}. 

\begin{table}[ht]
    \centering
    \begin{tabular}{|l|c|c|}
       \hline
       
         &  PPLN 1 & PPLN 2 \\ \hline
     SPDC spectra FWHM    & 91\,nm  & 92\,nm \\ \hline
     SHG conversion efficiency    & 34\,\%/W  & 39.2\,\%/W \\ \hline
     Brightness (pairs.s$^{-1}$.nm$^{-1}$.mW$^{-1}$) & $152\cdot10^6$ & $175\cdot 10^6$     \\ \hline 
         Efficiency & 6.2$\cdot10^{-8}$ & 7.16e-8  \\ \hline
         Coupling to SMF28 & 65.5\,\% &  62\,\%     \\ \hline
    \end{tabular}
    \caption{The main figures of merits have been characterized and reported here. It encompasses, the efficiency of up- and down-conversion, the spectral bandwidth of the emitted pairs but also the coupling efficiency of the photon pairs into single mode fiber.}   \label{tab:my_label}
\end{table}
From the measured SHG and SPDC efficiencies summarized above, the overall source brightness can be estimated as the number of photon pairs generated per unit of input telecom pump power in the bidirectionally pumped Sagnac configuration, yielding a total pair generation rate of 10.3 kpairs/s/nm/mW$^2$.

A short length single mode fiber at 780\,nm between PPLN 1 and 2 ensures more than 86\,dB of pump light attenuation but the circulator-filter only provides 47\,dB isolation from the input port. Therefore, when the entangled photons exit the Sagnac interferometer through the third port of the circulator-filter, additional pump rejection filtering is required. A cascaded set of two telecom add-drop DWDM at 1560.6\,nm (ITU21), 1562.23\,nm (ITU 19) and 1558.98\,nm (ITU 23), provides the required isolation and deterministically separates the photon pairs via spectral demultiplexing. In practice, the isolation of channels 20 and 22 from channel 21, using standard components, isn't high enough and have to be excluded from our multiplexing scheme. 100\,GHz-DWDMs are therefore used to select 20 symmetric ITU channel pairs around the pump wavelength, starting from ITU 19 and ITU 23 for most of the characterizations.

The general setup for the characterization is described as follow: the photon pairs are detected using superconducting nanowire single-photon detectors (IDQuantique) showing an efficiency of $80\%$ and a dark count rate of 50\,Hz. The individual detection rates and the analysis of the coincidence rate are computed using a time-to-digital converter (Swabian).

\section{Entanglement analysis}

\subsection{Energy-time encoding}
\begin{figure}[ht]
    \centering
    \includegraphics[width=1\linewidth]{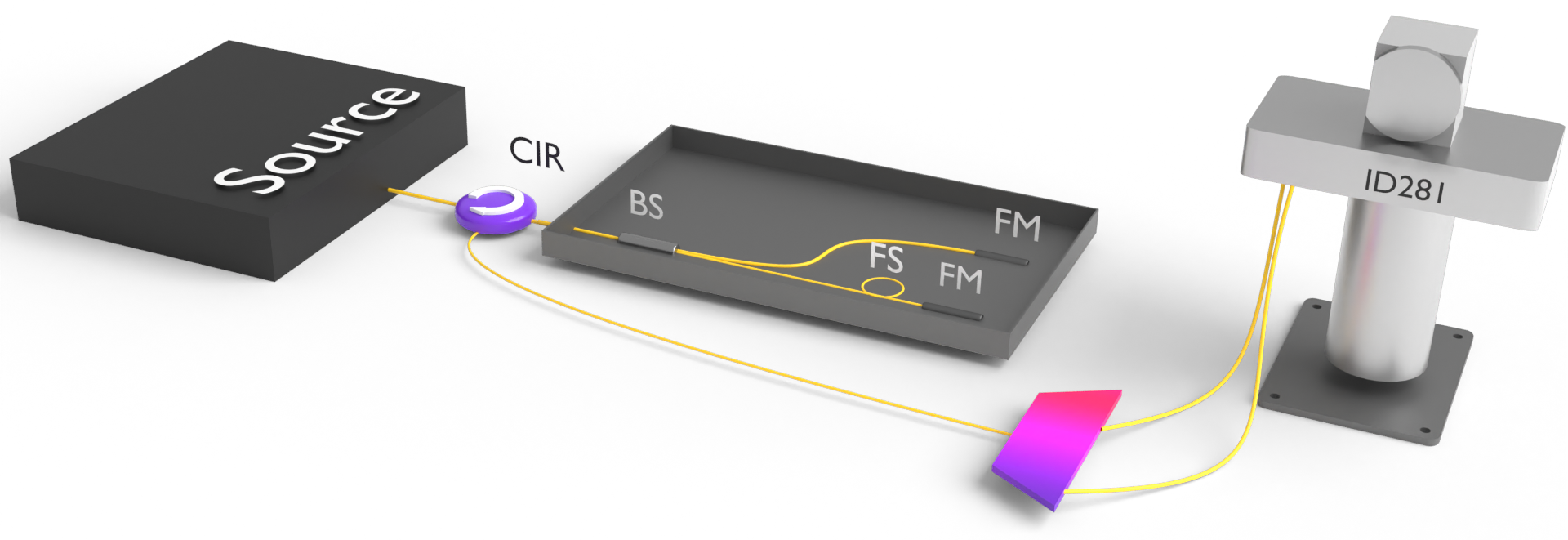}
            \includegraphics[width=1\linewidth]{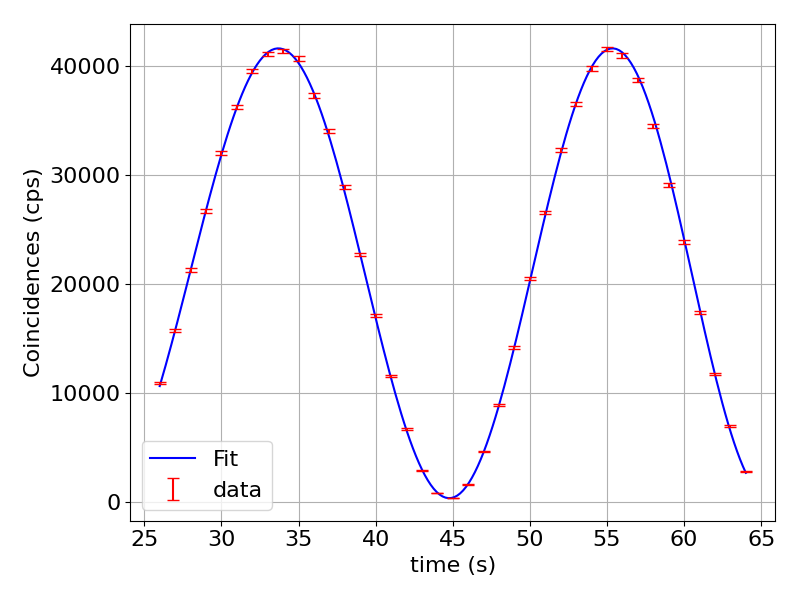}
    \caption{TOP : Fiber-based energy-time analyzer (Franson interferometer) used for two-photon interference measurements. BOTTOM : Two-photon interference fringes used to extract the raw energy-time visibility for the selected ITU channel pair.}
    \label{fig:ET_analyser}
\end{figure}

Energy-time entanglement is characterized using a passively stabilized fiber-based Franson interferometer in a Michelson configuration, as shown in \figurename~\ref{fig:ET_analyser}. The interferometer consists of a 50:50 fiber beam splitter and two Faraday mirrors, ensuring polarization-insensitive operation and perfect spatial mode overlap. Interference fringes are recorded by varying the relative phase between the interferometer arms using a piezoelectric fiber stretcher.
The interferometer FSR is chosen to satisfy the condition
\begin{equation}
        \Delta\nu_p < FSR < \Delta\nu_s\\
    \label{delay}
\end{equation} 
where the bandwidth of the individual photons $\Delta\nu_s=100$\,GHz and the photon pairs $\Delta\nu_p=1$\,kHz, respectively. In the present configuration, a practical FSR of  1\,GHz has been selected to be compatible with SNSPD timing jitter ($\sim$50\,ps).
 The raw visibility reaches $99 \pm 1\%$ on \figurename~\ref{fig:ET_analyser} for a coincidence detection rate of $\sim$500\,kHz and 500\,ms acquisition time, indicating high-quality energy-time entanglement for channels ITU 19 and ITU 23.

\subsection{Polarization encoding}
\begin{figure}[ht]
    \centering
    \includegraphics[width=1\linewidth]{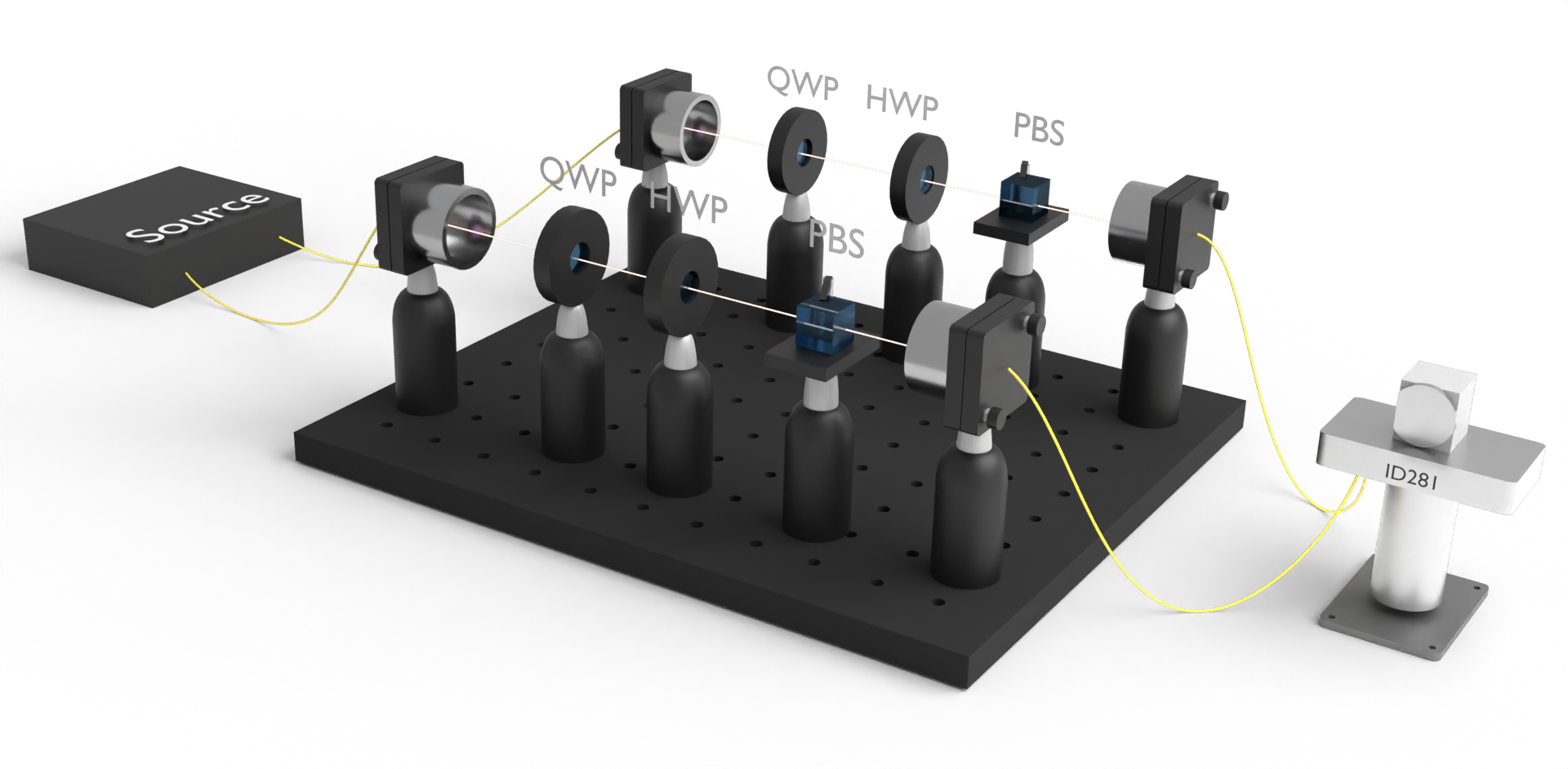}
    \caption{Polarization tomography module. A motorized set of quarter- and half-wave are used in front of polarizing beamsplitter in order to perform projective measurement of each photons along $|H\rangle$, $|V\rangle$, $|D\rangle$, and $|R\rangle$ quantum states. QWP: quarter waveplate, HWP: half waveplate, PBS: polarizing beamsplitter.}
    \label{fig:tomo_module}
\end{figure}
Polarization entanglement is characterized using a bulk polarization tomography module, schematically shown in \figurename~\ref{fig:tomo_module}. The analyzer consists of a sequence of quarter-wave and half-wave plates placed in front of a PBS. Bulk optics are employed to facilitate polarization manipulation and ensure stable projection bases. Fiber-coupled collimators are used to inject and collect photons at the input and output of the tomography module.

Complete reconstruction of the two-photon polarization density matrix is performed by measuring a set of polarization projections for each photon of the pair. Four polarization states, namely horizontal ($H$), vertical ($V$), diagonal ($D$), and right-circular ($R$), are analyzed for each photon, resulting in 16 projection combinations. Coincidence and singles rates are recorded for each projection pair and processed using the polarization tomography algorithm developed by Kwiat \textit{et al.}~\cite{dariano2003quantum}, yielding the reconstructed density matrix.
To evaluate the capability of the source for wavelength-multiplexed operation, polarization tomography has been performed on 20 symmetric ITU channel pairs spanning the C and L bands, as shown in \figurename~\ref{fig:raman}.
From the reconstructed density matrices, the fidelity with respect to the target Bell state $\ket{\Phi^+}$ and the state purity, defined as $F=\langle \Phi^+|\rho|\Phi^+\rangle$ and $P=\mathrm{Tr}(\rho^2)$, respectively, are extracted.

\begin{figure}[ht]
\centering
   \includegraphics[width=1\linewidth]{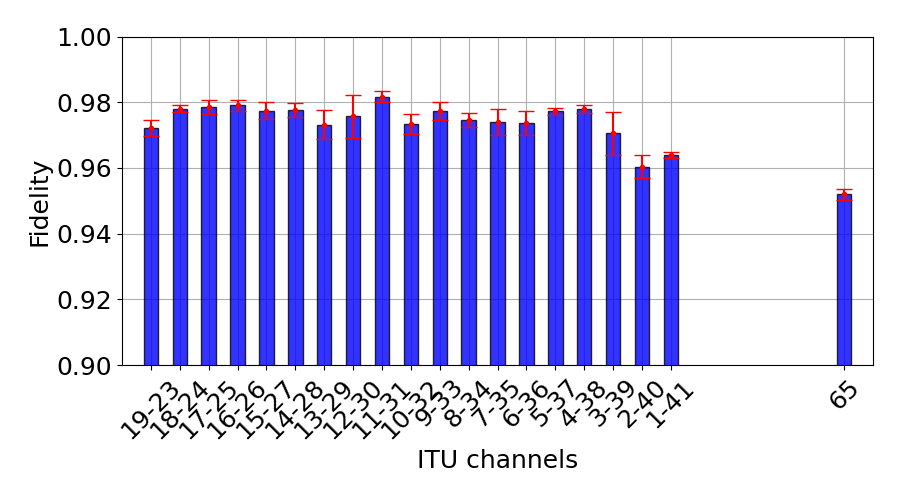}
      \includegraphics[width=1\linewidth]{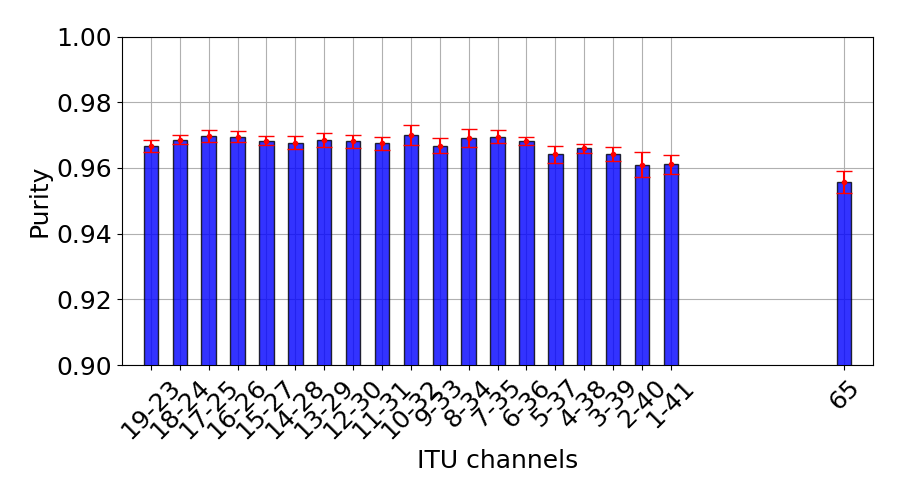}
   \caption{Raw fidelity and purity as function of the ITU channel pair. The coincidence and singles rates are recorded and most-likely density matrix calculated using~\cite{dariano2003quantum}. The fidelity to the state $\ket{\Phi^+}$ and the state purity are extracted from the density matrix.}
    \label{fig:fidelite}
\end{figure}

The tomography measurements have been automated and  all measurement parameters, including integration time, pump power, coincidence window, and calibration settings, are kept constant throughout the experiment. This approach is chosen to emulate realistic network operation, where the source simultaneously distributes entangled pairs over multiple wavelength channels without channel-specific optimization. Raw fidelities and purities obtained for all channel pairs are shown in \figurename~\ref{fig:fidelite}.

Raw fidelities and purities exceeding 96\% and 97\% are observed across the entire C band. Statistical uncertainties are primarily limited by Poissonian counting statistics, scaling as $\sqrt{N}$ due to the short integration time. A slight reduction of fidelity and purity is observed for the outermost channel pairs, which may be attributed to wavelength-dependent polarization effects along the fiber section to the tomography module and residual chromatic dispersion \cite{Dalidet_PRR2025}. Our automated measurement device relies on fixed projective measurements. A slight transformation of the polarization state entering the tomography module for the outermost channels modifies the effective measurement bases defined for the inner channels and leads to a reduction in the measured fidelity and purity, even though the underlying physical state remains unchanged.

However, the similar results for the 20 inner channels have an interesting consequence. It is worthnoting that, with 2$\times$20 single photon detectors available, one could perform a single quantum state tomography over all channels simultaneously with high fidelity. This broadband aspect of polarization entanglement envisions high speed entanglement distribution through wavelength multiplexing with a single analyzer associated with 20-multiplexed detectors at each location. In an polarization based QKD situation, similar to the one presented in section~\ref{QKD}, the gain factor over the secret key rate would be up to 20 times.

A comparison with related multiplexed Sagnac-based sources reported in the literature is limited by the scarcity of comparable experimental implementations. Nevertheless, similar entanglement quality has been reported in references~\cite{4chan_comparaison,freq_comb_sagnac}, despite differences in architecture and wavelength coverage.

\section{Practical characterization for real-field application}\label{QKD}

\begin{figure*}[ht]
    \centering
    \includegraphics[width=1\linewidth]{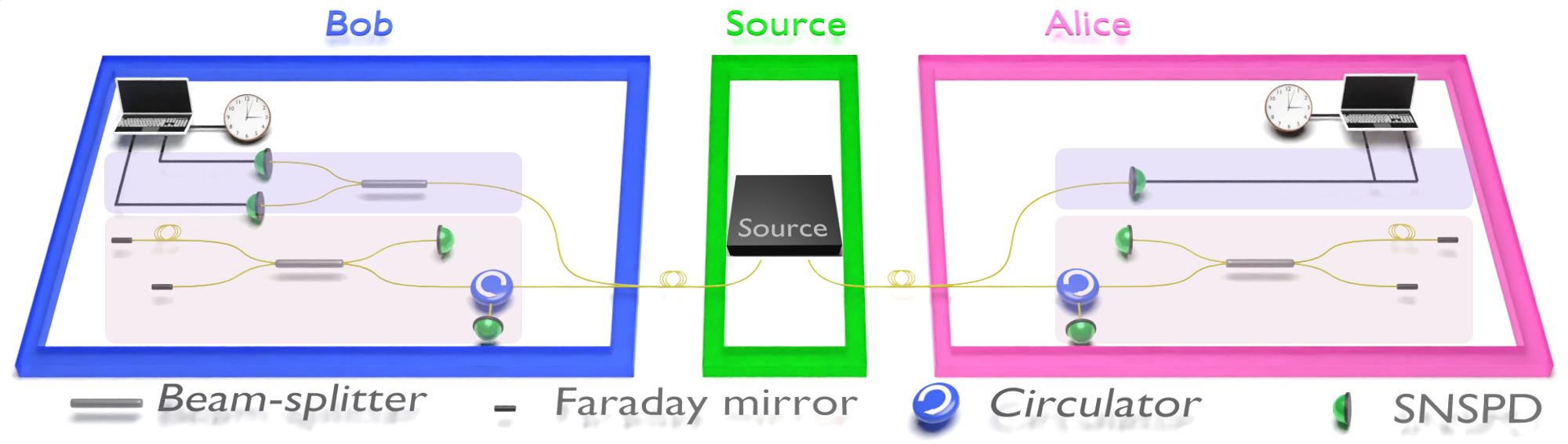}
    \caption{Deployed quantum-network testbed used for the energy-time QKD experiment. 
    At the center (green box), the source provide energy-time entanglement to Alice and Bob over ITU channels 19-23. Each analyzer (Alice and Bob) routes randomly the photons in either the Z or the X basis using a 50/50 beam splitter. The X basis is set up using two Franson interferometers, one at Alice’s station (pink box), the other at Bob’s (blue box). The Z basis consists of a 50/50 beam-splitter on Bob’s side with a short and a long path to the detectors while Alice only has one detector, short and long paths being created electronically. The protocol and setup have been extensively described in reference~\cite{yoyo,synchro_pelet}.}
    \label{fig:reseau}
\end{figure*}

\begin{figure}[ht]
         \centering
         \includegraphics[width=1\linewidth]{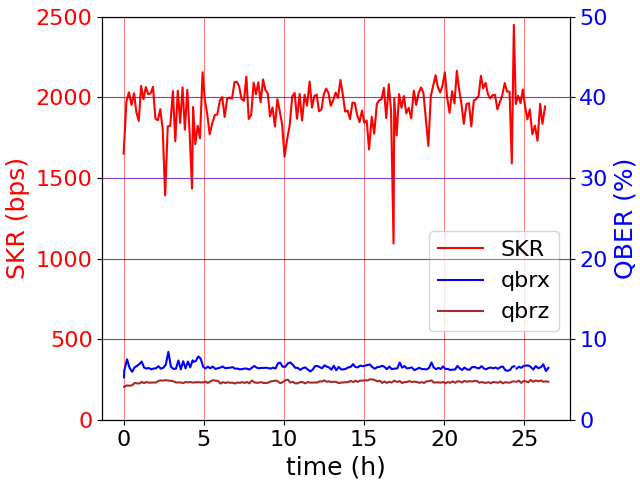}
         \caption{Secret key rate and its associated QBERs in the X and Z bases monitored during 27\,h and over 50\,km link. SKR and QBERs are calculated on the fly.}
         \label{fig:skr}
     \end{figure}

To further assess the operational robustness of the source under realistic conditions, an energy-time quantum key distribution (QKD) experiment is carried out on the deployed quantum network available for our laboratory~\cite{yoyo}. This network, originally developed and optimized for energy-time entanglement distribution, provides a relevant testbed to evaluate the behavior of the source in a practical multi-user environment. It enables the validation of the long-term stability and resilience of the source when deployed in a harsh environment.
The experimental setup consists of the Sagnac source interfaced with the existing network infrastructure, including standard telecom fibers, actively stabilized unbalanced Franson interferometers used as analyzers, and single-photon detectors operating at telecom wavelengths, as shown in \figurename~\ref{fig:reseau}. The measurements have been performed on ITU channel pair 19-23. All system parameters are kept fixed throughout the experiment, and no realignment or recalibration of the source is applied during the measurement campaign.
Secret key rates (SKR) and quantum bit error rates (QBER) are continuously monitored over extended periods, providing direct insight into the temporal stability of the source. The measured mean SKR of 1.95 kbps and QBER values of 6.5\% and 4.7\% in the $X$ and $Z$ bases respectively, have been measured for more than 26\,h, as shown on \figurename~\ref{fig:skr}.
Occasional abrupt drops in the secret key rate are observed during the acquisition. The pronounced drops (e.g. at 2\,h, 4\,h, and 16\,h) are due to latency in real-time post-processing.
The X-basis QBER is derived from the Franson interferometers visibility. Their relative phase is actively stabilized in real time to minimize the QBER$_X$. The peaks observed in \figurename~\ref{fig:skr} result from random and sudden phase drifts affecting one of the interferometers. In contrast, the QBER in the $Z$ basis is determined during the error correction stage and corresponds to the fraction of errors corrected in the raw key.

\section{Conclusion}

A fully fibered Sagnac-interferometer source of entangled photon pairs operating at telecom wavelengths has been presented and thoroughly characterized. Relying exclusively on standard fiber-optic components and nonlinear PPLN waveguides, the proposed architecture combines compactness, mechanical stability, and ease of use with high entanglement quality. The source supports both polarization and energy-time entanglement without modification of the generation stage, enabling versatile operation depending on the targeted application.

Detailed entanglement characterization demonstrates high performance in both encoding schemes. Polarization entanglement is quantified through full state tomography over 20 symmetric ITU channel pairs spanning the C and L bands, yielding raw fidelities and purities exceeding 96\% and 97\%, respectively. These results confirm the suitability of the source for wavelength-multiplexed operation and multi-channel quantum networking. Complementary energy-time measurements based on two-photon interference further demonstrate high visibility, about $99.2\%\pm1.5\%$.

Beyond laboratory characterization, the robustness of the source is validated through long-duration operation in a deployed quantum network. An energy-time QKD experiment is used as a system-level test of stability. The measured secret key rates and QBER values remain stable over extended periods and are in good agreement with numerical simulations based on independently measured system parameters. Performance limitations are primarily attributed to the additional optical losses associated with the dual-encoding design, rather than to any degradation of entanglement quality.

Overall, these results demonstrate that the proposed Sagnac-based source constitutes a mature and resilient building block for practical quantum communication systems. Its intrinsic stability, broadband operation, and compatibility with wavelength multiplexing make it particularly well suited for plug-and-play quantum networking architectures. Future work will focus on the implementation of polarization-based QKD using the same source, providing a stringent test of long-term phase stability and polarization entanglement preservation under realistic and potentially harsh environmental conditions.

\begin{acknowledgments}
This work has been conducted within the framework of the French government financial support managed by the Agence Nationale de la Recherche (ANR), within its Investments for the Future program, under the Université Côte d'Azur UCA-JEDI project (ANR-15-IDEX-01), under the SOLUQs project (ANR-20-ASTQ-0003), and under the Stratégie Nationale Quantique through the PEPR QCOMMTESTBED project (ANR 22-PETQ-0011). This work has also been conducted within the framework of the OPTIMAL project, funded by the European Union and the Conseil Régional SUD-PACA by means of the “Fonds Européens de développement regional” (FEDER). The authors also acknowledge financial support from the European Commission, through the Project 101114043-QSNP and 101091675-FranceQCI. The authors are grateful to S. Canard, A. Ouorou, L. Chotard, L. Londeix, and to Orange for the installation our quantum network. The authors also thank the Métropole Nice Côte d'Azur and the Inria Centre at Université Côte d'Azur for the access to their buildings and Thales Alenia Space and the LiP6 for their scientific support. The authors also acknowledge IDQuantique, Exail and Swabian Instruments GmbH teams for the technical support.
\end{acknowledgments}


\end{document}